\documentclass{article}

\PassOptionsToPackage{square, numbers}{natbib}
\usepackage[preprint]{neurips_2021}

\usepackage[utf8]{inputenc} 
\usepackage[T1]{fontenc}    
\usepackage{hyperref}       
\usepackage{url}            
\usepackage{booktabs}       
\usepackage{amsfonts}       
\usepackage{nicefrac}       
\usepackage{microtype}      
\usepackage{xcolor}         
\usepackage{mathtools}
\usepackage{lineno}
\usepackage{import}
\usepackage{amsmath}
\usepackage{amsfonts}
\usepackage{rotating}
\usepackage[skip=10pt]{caption}

\DeclarePairedDelimiter\floor{\lfloor}{\rfloor}

\title{How informative is the Order Book Beyond the Best Levels? Machine Learning Perspective}

\author{%
  Dat Thanh Tran\\
  Department of Computing Sciences\\
  Tampere University\\
  Tampere, Finland \\
  \texttt{thanh.tran@tuni.fi} \\
  \And
  Juho Kanniainen \\
  Department of Computing Sciences \\
  Tampere University\\
  Tampere, Finland \\
  \texttt{juho.kanniainen@tuni.fi} \\
  \AND
  Alexandros Iosifidis\\
  Department of Engineering\\
  Aarhus University\\
  Aarhus, Denmark \\
  \texttt{ai@ece.au.dk} \\
}

\begin{document}

\maketitle

\begin{abstract}
Research on limit order book markets has been rapidly growing and nowadays high-frequency full order book data is widely available for researchers and practitioners. However, it is common that research papers use the best level data only, which motivates us to ask whether the exclusion of the quotes deeper in the book over multiple price levels causes performance degradation. In this paper, we address this question by using modern Machine Learning (ML) techniques to predict mid-price movements without assuming that limit order book markets represent a linear system.  We provide a number of results that are robust across ML prediction models, feature selection algorithms, data sets, and prediction horizons. We find that the best bid and ask levels are systematically identified not only as the most informative levels in the order books, but also to carry most of the information needed for good prediction performance. On the other hand, even if the top-of-the-book levels contain most of the relevant information, to maximize models' performance one should use all data across all the levels. Additionally, the informativeness of the order book levels clearly decreases from the first to the fourth level while the rest of the levels are approximately equally important.
\end{abstract}

\section{Introduction}

The modern stock markets are mainly operated by the electronic limit order book (LOB) mechanism, which allows traders and algorithms to immediately use order-book data beyond the best levels for their trading decisions. Intuitively, the contents of a LOB can be informative and useful for trading purposes for many reasons. The order-book asymmetry may reflect trader sentiment or the presence of well informed traders, for which reason one might want to trade in front of its heavy side \citep{harris2005information}. Moreover, data beyond the best levels reflects the order-book liquidity, i.e. the quantity immediately available for trading and therefore the price of immediacy. Available liquidity beyond the best levels matters especially when investors' impatience generates a sudden liquidity demand across multiple price levels, leading to large movements in stock prices \citep{siikanen2017limit, siikanen2017drives}.

Even if the data beyond the best levels were informative, there can be many reasons to use the best-level data only, often referred to as Level-I  data, to answer certain research questions. Firstly, parsimonious and tractable models that accomplishes a desired level of explanation or prediction with as few predictor variables as possible can be preferred for the sake interpretability. If models are large and complex, the parameter estimates can be shaky (having large standard errors), leading to statistical insignificance and thus uncertain conclusions.
Secondly, Level-I data is easier to obtain than Level-II data, which includes quotes deeper in the book over multiple price levels. It is usually the case that Level-I data is well available for academic research. However, when it comes to the maximization of model's predictive power, the current literature lacks answers to the question of how informative the top quotes are compared the quotes beyond them. Moreover, it is unclear whether one should use multi-level order book data, even at the cost of additional complexity of the model.

Seeking answers to the research questions above is important in many ways. A major portion of the extant financial and econometric literature rely on Level-I data for different research questions. Thus, it is important to understand the impact on the results when excluding the information beyond the best level. In literature, Level-I  data has been used for various purposes, for example, to analyze micro-structure noise \citep{ait2011ultra,bandi2006separating}, price impact \citep{dufour2000time,engle2004impacts,bouchaud2004fluctuations,eisler2012price}, optimal trading strategies \citep{guilbaud2013optimal}, algorithmic trading \citep{hendershott2011does,chaboud2014rise}, price prediction or order-book dynamics modeling \citep{cont2010stochastic,cont2013price}. Even if this paper focuses on the last topic, i.e. the modeling and prediction of the stock prices with order book data, we believe that our results also shed light on other topics since we are assessing the informational content of the empirical LOB data in terms of the price formation, which is related to trading strategies as well as the use of algorithmic trading. If the inclusion of data beyond the best level does not lead to noticeable improvements of a prediction model, then it is sufficient to solely rely on Level-I data, which in turn validate the methodological and data choices of existing works relying only on Level-I data.

On the other hand, there is also a branch of literature that uses multi-level order-book data to analyze the use of the limit versus market orders \citep{anand2005empirical,linnainmaa2010limit}, order book liquidity \citep{pardo2012hidden,siikanen2017limit,siikanen2017drives}, market impact \citep{farmer2005predictive,hautsch2012market}, and price formation \citep{abergel2013mathematical}. Recently, the use of complete order-book data has become popular in predicting order-book dynamics or price movements with advanced machine learning techniques, particularly in the quantitative finance and machine learning literature (see, for example \cite{dixon2017classification,tran2018temporal,sirignano2019universal,sirignano2019deep,zhang2019deeplob,makinen2019forecasting,nousi2019machine,passalis2019deep} and references therein). Moreover, recently Deep Reinforecement Learning has been used for high-frequency trading with LOB data \citep{briola2021deep}. Even if modern deep-learning techniques allow a huge number of input variables (as long as the size of the data is large), using irrelevant input variables is not only unnecessary but can also reduce the robustness of the models. Irrelevant data, which potentially acts as a source of noise, can adversely affect the accuracy of a model that captures the underlying relationship between the variables. Moreover, the extra computational burden introduced by additional information can be significant, making the analysis impractical for high thoughput applications. For these reason, it is important to know if complete Level-II order book data indeed maximizes the predictive power of machine learning models. If this is the case, then one could argue that it is better to use data-driven machine learning techniques with all the available data than parsimonious and tractable models that rely only on the top quotes.

In fact, the question about the importance of the order book data beyond the best level has been barely addressed in the literature. While there are existing papers that have considered the question about informational contents of LOBs, such as \citep{pascual2003pieces,cao2008order,cao2009information,duong2014anonymity}, they rely on such simple linear models. In reality, the way how investors and algorithms use order-book data for their trading decisions can be very complicated and vary over different circumstances. If one estimates a specific model assuming pre-specified (linear) relations between the explanatory and dependent variables, the drawn conclusions can be misleading because of the possibility of model misspecification. More generally, in variable selection (feature selection), it is crucial not to restrict the model to be linear, and in fact, to have any restriction that can incorrectly to falsify the importance of a variable. For that reason, the assumption of linearity is not acceptable when it comes to truly non-linear systems, what stock markets definitely represent \citep{hsieh1991chaos}.

In this paper, instead of relying on the linear model paradigm, we aim to address the question about the informativeness of the order book data beyond the best levels by allowing the involved variables to interact in a complex, nonlinear manner. To our best knowledge, this is the first paper that addresses this research question by a data-driven approach in which the necessity of each explanatory variable and their complex, nonlinear interaction (which also encapsulates the linear case) are jointly estimated. More specifically, we employ state-of-the-arts modeling tools from the machine learning community in our method, namely deep neural networks. It has been shown that feed-forward neural networks with bounded activation functions or rectified linear units are universal function approximators \citep{chen2018universal}. Simply put, a universal function approximator is  guaranteed to have the capacity to approximate arbitrarily well any function that generates the observed data. Thus, from the theoretical perspective, deep neural networks are suitable tools to model the underlying complex relationship between the limit order information and any dependent targets since the only assumption required is the existence of such a relationship. Besides, the most prominent empirical successes over the last few years in financial applications such as stock prediction \cite{long2019deep, fu2021mhier, gunduz2017intraday, zhou2016financial}, portfolio selection and optimization \citep{cao2020delafo, zhang2020deep}, factor and risk analysis \citep{addo2018credit, leo2019machine}, derivatives hedging \citep{cao2019deep, du2020deep} and so on, are mainly driven by deep neural network solutions.

In order to discover the importance of different explanatory variables with respect to the dependent variables of interest, our work adapts and integrates well-studied automatic feature selection methods into the training procedure of deep neural networks. By doing so, we are able to gauge the the contribution of explanatory variables to the final forecasting decision in a complete data-driven manner. As the name implied, feature selection \citep{chandrashekar2014survey} aims to select the most relevant subset of the explanatory variables to achieve similar or even better learning performance than using all explanatory variables. By developing feature selection methods, not only are we able to identify the most relevant information from the LOB but also to reduce the computational inference cost when deploying the estimated models in production.

The machine learning literature mainly concerns about learning models and the final performances. At the same time, we are not aware of any prior work that investigates the consistency among the solutions obtained from a feature selection method applied to different backbone estimation models or even across the solutions obtained from different feature selection methods. Because our main focus is on the informational content of the LOB, interpretations of the outcome are only valid when there is a consensus among different feature selection methods using different backbone estimation models, estimated with observed data from different markets. For this reason, in our work, we adapt and integrate two popular feature selection methods, namely \textit{Backward Elimination} (\texttt{BE}) \citep{efroymson1960multiple, hocking1976biometrics} and \textit{Binary Particle Swarm Optimization} (\texttt{BPSO}) \citep{kennedy1997discrete}, into two state-of-the-arts neural network models that are specifically designed for predicting the mid-price movement in LOBs, namely \textit{Deep Convolutional Neural Network for LOB } (\texttt{DeepLOB}) \citep{zhang2019deeplob} and \textit{Temporal Attention Bilinear Layer Network} (\texttt{TABL}) \citep{tran2018temporal}.

With extensive experimentation using data from two different markets (US and Nordic), we find out that there is indeed a consensus between different combinations of (i) neural network models, (ii) feature discovery methods, and (iii) markets: the top level of the LOBs provides the most important source of information in predicting the future movements of the mid-price. In addition, there is also a consensus in the ranking between the top three levels of the order-book in terms of importance: the top three levels are also ranked in the same order as their execution priority in the order-book. Our analysis also points out that orders beyond the best level indeed provide complementary information in the prediction of mid-price movements, accounting for two to three percents of performance improvements compared to the cases where only the most important quotes are used.

\section{Informational Content Discovery via Feature Selection}\label{main-method}

In order to discover the predictive power of different levels in the LOB, we selected two feature selection methods, namely Backward Elimination (\texttt{BE}) \citep{efroymson1960multiple, hocking1976biometrics} and Binary Particle Swam Optimization (\texttt{BPSO}) \citep{kennedy1997discrete} with different search trajectories. \textit{Here, the $k$-th level in the LOB refers to the price and volume of the $k$-th best bid and best ask orders}. While \texttt{BE} produces a candidate subset of a given size that may achieve the best performance, the search trajectory of \texttt{BPSO} can travel the entire search space and converges at a subset of any size, which is not predefined. Since our selected classifiers (\texttt{TABL} and \texttt{DeepLOB}) are estimated using stochastic gradient descent, it is possible that different runs of either \texttt{BE} or \texttt{BPSO} can generate different subsets. Thus, to analyze the informational content of our inputs, we look for any consensus among the solutions provided by both methods applied to the selected classifiers, using two different markets' data. Our approach to look for such consensus is elaborated in Section \ref{experiment}.

The idea of \texttt{BE} is very simple: the algorithm starts with the set of all input features, then gradually identifies and removes the most irrelevant feature, one at a time, until there is only a single level left or the learning performance falls below a pre-defined threshold. The majority of existing literature that uses \texttt{BE} employed simple linear models, thus $F$-test or $t$-test are popular criteria to identify the irrelevant feature at each elimination step. For the approach that utilizes a linear regressor, we only need to perform one estimation of the model's parameters and use them to identify which feature to remove. In our case where highly nonlinear neural networks are employed, there is no straight-forward method to quantify the importance of a given input variable based on the optimized weights. For this reason, in order to determine data from which order-book level should be discarded in an elimination round having $K$ remaining levels, we train $K$ different neural network instances, each of which corresponds to leaving out the quotes of a potentially irrelevant level. The irrelevant level is then identified with the network instance that achieves the lowest prediction performance. In our experiments, we performed the elimination routine until there was only one level left. 

While \texttt{BE} resembles more of a series of heuristic tests to identify important features, \texttt{BPSO} is a generic, well-studied meta-heuristic optimizer for binary variables, rather than a dedicated feature selection method. That is, \texttt{BPSO} optimizes a binary vector such that a fitness function computed on it is maximized. In order to use \texttt{BPSO} to perform feature selection, one would define a binary mask vector having the same dimension as the number of features, which represents the selection of a particular subset of features. The values of this binary mask vector are randomly initialized and iteratively updated until convergence. \texttt{BPSO} applied in such manner considers each input variable independently while in our case, we want to form subset of the quote values that correspond to specific order-book levels. This necessitates modifications as how \texttt{BPSO} is formulated for LOBs.

The inputs to the neural network classifiers are the most recent order-book events, which include the top ten bid prices ($p^b_1(t), \dots, p^b_{10}(t)$) and volumes ($v^b_1(t), \dots, v^b_{10}(t)$), and top ten ask prices ($p^a_1(t), \dots, p^a_{10}(t)$) and volumes ($v^a_1(t), \dots, v^a_{10}(t)$), with $t$ indicating the time index. These variables are organized in the following matrix to reflect a multivariate time-series nature of the order-book events:

\begin{equation}\label{eq12}
\mathbf{X}_t = \begin{bmatrix}
	p^a_1(t-T-1) &  \dots & p^a_1(t) \\
	v^a_1(t-T-1) &  \dots & v^a_1(t) \\
	p^b_1(t-T-1) &  \dots & p^b_1(t) \\
	v^b_1(t-T-1) &  \dots & v^b_1(t) \\
	\vdots 		 & \dots  & \vdots \\
	p^a_{10}(t-T-1) &  \dots & p^a_{10}(t) \\
	v^a_{10}(t-T-1) &  \dots & v^a_{10}(t) \\
	p^b_{10}(t-T-1) &  \dots & p^b_{10}(t) \\
	v^b_{10}(t-T-1) &  \dots & v^b_{10}(t) \\
\end{bmatrix} \in \mathbb{R}^{40 \times T},
\end{equation}
where $T$ denotes the number of the most recent order-book events that a neural network predictor uses as inputs. In addition, each sample $\mathbf{X}_t$ comes with a label $y_t$ that indicates the future mid-price movement.

Let us denote by $\mathbf{s} \in \mathbb{R}^{10}$ the binary vector that indicates a particular subset of ten levels, with $\mathbf{s}[i] = 1$ indicating the inclusion of level $i$ and $\mathbf{s}[i] = 0$ otherwise. In addition, we also denote by $\mathbf{M}(\mathbf{s}) \in \mathbb{R}^{40\times T}$ a binary matrix that is formed from $\mathbf{s}$ as
\begin{gather}\label{eq13}
\mathbf{M}(\mathbf{s})[i, j] = \mathbf{s}[k], \quad \forall j=1, \dots, T \\
k = \floor*{(i-1)/4} + 1, \quad \forall i=1, \dots, 40.
\end{gather}
That is, $\mathbf{M}(\mathbf{s})$ denotes the binary matrix that can be used to mask out all the values in $\mathbf{X}_t$ that belongs to unchosen levels expressed in $\mathbf{s}$. Here we should note that $\mathbf{M}(\mathbf{s})$ can also be defined via a matrix multiplication operation as $\mathbf{M}(\mathbf{s}) = \mathbf{I}_1 \mathbf{s} \mathbf{I}_2$ with $\mathbf{I}_1 \in \mathbb{R}^{40 \times 10}$ and $\mathbf{I}_2 \in \mathbb{R}^{1 \times T}$ being two constant matrices with appropriate ones and zeros. Now the optimization objective of \texttt{BPSO} can be stated as
\begin{equation}\label{eq14}
	\underset{\mathbf{s}}{\mathrm{argmax}} \quad \mathcal{F}_1\big(\{y_t, \mathcal{N}(\mathbf{X}_t \odot \mathbf{M}(\mathbf{s})) | t=1, \dots\}\big),
\end{equation}
where $\mathcal{N}$ denotes the function representing the neural network giving predictions. $\mathcal{F}_1$ denotes the function that computes average F1 measure in a classification problem given the set of true labels and predicted labels. The element-wise multiplication operation at the end of Eq. (\ref{eq14}) is denoted as $\odot$.

Here we should note that the F1 measure is used in our analysis because it reflects the trade-off between precision and recall, which is especially important for problems with skewed distributions as observed in our data.

The idea in \texttt{BPSO} is that we may discover an optimal solution faster by using a list of candidate solutions, which are also called particles. Let us denote the $k$-th particle in the swarm as $\mathbf{s}_k$ and the size of the swarm as $K$. The swarm of particles moves together in the search space to find the optimal solution, thus, each of them has a velocity, which is denoted as $\mathbf{v}_k \in \mathbb{R}^{10}$. At each movement, a particle moves to a new position with its velocity, which is computed based on its past experience and also the experiences of other particles. Let us denote the best position that the $k$-th particle has visited as $\mathbf{s}_k^{*}$, and the best position that the entire swarm has visited as $\mathbf{s}^{*}$. The $k$-th particle adjusts its velocity with the following rule:

\begin{equation}
	\mathbf{v}_k^{(i+1)} = w \odot \mathbf{v}_k^{(i)} + c_1 \cdot r_1 \cdot (\mathbf{s}_k^{*} - \mathbf{s}_k^{(i)}) + c_2 \cdot r_2 \cdot (\mathbf{s}^* - \mathbf{s}_k^{(i)}) \label{eq15.1},
\end{equation}
where $i$ represents the iteration index in the optimization routine. $w$ is the inertia term, which is a hyperparameter of the method to control the impact of the old velocity on the new one. The value of $w$ is often reduced gradually as the swarm progresses. $c_1 \in \mathbb{R}$ and $c_2 \in \mathbb{R}$ are acceleration constants and $r_1$ and $r_2$ are randomly sampled scalars from the uniform distribution $[0, 1]$. To control the maximum velocity of a particle at each iteration, the velocity is limited within a range $[-\mathbf{v}_{\textrm{max}}, \mathbf{v}_{\textrm{max}}]$.

Since a particle can only move in a discrete space of 0 and 1, the velocity is used to determine the probability that a particle will move to position 0 or 1. With the new velocity $\mathbf{v}_k^{(i+1)}$, the $k$-th particle updates its position using the following rule:

\begin{equation}\label{eq15.2}
	\mathbf{s}_k^{(i+1)}[j] = \begin{dcases} 1 \quad \textrm{if } r_3 < \textsf{sigmoid}\left(\mathbf{v}_k^{(i+1)}[j] \right)\\
				0 \quad \textrm{else},
	\end{dcases}
\end{equation}
where $\mathbf{s}_k^{(i+1)}[j]$ and $\mathbf{v}_k^{(i+1)}[j]$ denote the $j$-th element in $\mathbf{s}_k^{(i+1)}$ and $\mathbf{v}_k^{(i+1)}$, respectively. $r_3$ is a randomly sampled scalars from the uniform distribution $[0, 1]$.

Given the new positions of the swarm, \texttt{BPSO} evaluates the fitness of each particle by optimizing the neural network's parameters, i.e., for the $k$-th particle, we need to solve the following optimization problem with stochastic gradient descent:
\begin{equation}\label{eq16}
	\Theta^* = \underset{\Theta}{\mathrm{argmin}} \sum_{t} L\left(y_t, \mathcal{N}\left(\mathbf{X}_t \odot \mathbf{M}\left(\mathbf{s}_k^{(i)}\right)\right)\right),
\end{equation}
where $\Theta$ is the parameters of the neural network $\mathcal{N}$, and $\Theta^*$ denotes its optimal values obtained from stochastic gradient descent. $L$ denotes the cross-entropy loss function.

The fitness for the $k$-th particle at the $i$-t iteration is defined as the F1 score obtained at $\Theta^*$. Based on new fitness values for all particles, \texttt{BPSO} checks and updates $\mathbf{s}_k^*$ and $\mathbf{s}^*$ if new best solutions emerge. The algorithm repeats the aforementioned procedure for a pre-defined number of iterations.

\section{Empirical Results}\label{experiment}

We conducted our experiments using the LOB data coming from two different markets: the Nordic market and the US market. For the Nordic market, we collected orders from five companies that are listed on Helsinki Stock Exchange, namely Kesko, Outokumpu, Rautaruukki, Sampo and Wartsila, and for the US market, we used orders coming from Amazon and Google. All of our data is procured from TotalView-ITCH feed covering a period of ten working days from the 22nd of September 2015 to the 5th of October 2015. This resulted in approximately 13 millions events for the US dataset and 4 millions events for the Nordic dataset. For each dataset, the data of the first seven days was used to to train and validate the models and the data covering the last three days was used as the held-out test set. For each market, we aggregated the data from all stocks and used them to train stock-agnostic models to predict the future movement of the mid-price after $H = \{10, 20, 50\}$ order events. The future movements of the mid-price belongs to one of the three categories: up, down, and stationary. Given two types of prediction networks (\texttt{TABL} and \texttt{DeepLOB}), two feature selection methods (\texttt{BE} and \texttt{BPSO}), two datasets (US and Nordic), and three prediction horizons ($H = \{10, 20, 50\}$ events), our total experiment set consists of $24$ different configurations. For each configuration, we repeated the experiment $20$ times and reported the mean and standard deviation values for F1 score measured on the held-out test set. The supplementary material provides additional information regarding the labeling process of the mid-price movements, the preprocessing steps, the hyperparameter values and other optimization details.

\subsection{Are the top quotes the most important predictors?}

Based on the experiment results collected, the first question that we sought an answer is whether the assumption about the top quotes being the most informative information of the LOBs still holds even when using highly nonlinear models for the modeling task. Let us recall from Section \ref{main-method} that as the elimination goes in \texttt{BE}, we sequentially obtain the most important subsets of LOB levels of decreasing sizes, and also the associated test performance that can be achieved from the selected subset. This process was conducted until there was only one level remaining. If we look into the percentage that a given level appeared as the only remaining level after running \texttt{BE}, out of $20$ experiment repetitions for each configuration, we obtained the following statistics for ten levels in Table \ref{t1}.

\begin{table*}[h]
\begin{center}
	\caption{The percentage that a given level  (on the both sides of the book) appeared as the only remaining level when using \texttt{BE}.}\label{t1}
	\resizebox{1.0\textwidth}{!}{
	%
\begin{tabular}{lccccccccccccccc}
\cmidrule{2-16}      & \multicolumn{7}{c}{US Data}                           &       & \multicolumn{7}{c}{Nordic Data} \\
\cmidrule{2-8}\cmidrule{10-16}      & \multicolumn{3}{c}{DeepLOB} &       & \multicolumn{3}{c}{TABL} &       & \multicolumn{3}{c}{DeepLOB} &       & \multicolumn{3}{c}{TABL} \\
\cmidrule{2-4}\cmidrule{6-8}\cmidrule{10-12}\cmidrule{14-16}$H$     & 10    & 20    & 50    &       & 10    & 20    & 50    &       & 10    & 20    & 50    &       & 10    & 20    & 50 \\
\cmidrule{1-4}\cmidrule{6-8}\cmidrule{10-12}\cmidrule{14-16}Level 1 & 100\% & 100\% & 100\% &       & 100\% & 100\% & 100\% &       & 95\%  & 90\%  & 95\%  &       & 65\%  & 95\%  & 100\% \\
Level 2 & 0\%   & 0\%   & 0\%   &       & 0\%   & 0\%   & 0\%   &       & 0\%   & 5\%   & 0\%   &       & 25\%  & 5\%   & 0\% \\
Level 3 & 0\%   & 0\%   & 0\%   &       & 0\%   & 0\%   & 0\%   &       & 0\%   & 0\%   & 0\%   &       & 5\%   & 0\%   & 0\% \\
Level 4 & 0\%   & 0\%   & 0\%   &       & 0\%   & 0\%   & 0\%   &       & 0\%   & 5\%   & 5\%   &       & 5\%   & 0\%   & 0\% \\
Level 5 & 0\%   & 0\%   & 0\%   &       & 0\%   & 0\%   & 0\%   &       & 0\%   & 0\%   & 0\%   &       & 0\%   & 0\%   & 0\% \\
Level 6 & 0\%   & 0\%   & 0\%   &       & 0\%   & 0\%   & 0\%   &       & 0\%   & 0\%   & 0\%   &       & 0\%   & 0\%   & 0\% \\
Level 7 & 0\%   & 0\%   & 0\%   &       & 0\%   & 0\%   & 0\%   &       & 5\%   & 0\%   & 0\%   &       & 0\%   & 0\%   & 0\% \\
Level 8 & 0\%   & 0\%   & 0\%   &       & 0\%   & 0\%   & 0\%   &       & 0\%   & 0\%   & 0\%   &       & 0\%   & 0\%   & 0\% \\
Level 9 & 0\%   & 0\%   & 0\%   &       & 0\%   & 0\%   & 0\%   &       & 0\%   & 0\%   & 0\%   &       & 0\%   & 0\%   & 0\% \\
Level 10 & 0\%   & 0\%   & 0\%   &       & 0\%   & 0\%   & 0\%   &       & 0\%   & 0\%   & 0\%   &       & 0\%   & 0\%   & 0\% \\
\bottomrule
\end{tabular}%

	}
\end{center}
\end{table*}

As it is obvious from Table \ref{t1}, the first LOB level consistently appeared with high percentages as the last level remaining during the elimination in \texttt{BE} across different configurations. Most notably, for the US data, the results of all configurations unanimously selected level 1 as the most predictive level. The results for the Nordic data also indicated the same phenomenon with level 1 being selected at least $18$ out of $20$ times (90\%) for all configurations, with the only exception when using \texttt{TABL} to predict at $H=10$. We will now look into the results obtained from \texttt{BPSO} from the same aspect. Table \ref{t2} shows the percentage that a given level appeared in the final solution produced by \texttt{BPSO}, out of $20$ experiment runs for each configuration. Even though we observed more variation in the results (mainly associated with the Nordic data) when running \texttt{BPSO}, the first level was repeatedly selected by \texttt{BPSO} in its solution with percentages exceeding other levels by large margins. Table \ref{t1} and Table \ref{t2} show that there is indeed a consensus about the first level being the most important predictor.

\begin{table*}[h]
\begin{center}
	\caption{The percentage that a given level (on the both sides of the book) appeared in the final solution of \texttt{BPSO}.}\label{t2}
	\resizebox{1.0\textwidth}{!}{
	%
\begin{tabular}{lccccccccccccccc}
\cmidrule{2-16}      & \multicolumn{7}{c}{US Data}                           &       & \multicolumn{7}{c}{Nordic Data} \\
\cmidrule{2-8}\cmidrule{10-16}      & \multicolumn{3}{c}{DeepLOB} &       & \multicolumn{3}{c}{TABL} &       & \multicolumn{3}{c}{DeepLOB} &       & \multicolumn{3}{c}{TABL} \\
\cmidrule{2-4}\cmidrule{6-8}\cmidrule{10-12}\cmidrule{14-16}$H$     & 10    & 20    & 50    &       & 10    & 20    & 50    &       & 10    & 20    & 50    &       & 10    & 20    & 50 \\
\cmidrule{1-4}\cmidrule{6-8}\cmidrule{10-12}\cmidrule{14-16}Level 1 & 100\% & 100\% & 60\%  &       & 100\% & 100\% & 80\%  &       & 60\%  & 50\%  & 50\%  &       & 50\%  & 90\%  & 70\% \\
Level 2 & 20\%  & 10\%  & 30\%  &       & 0\%   & 10\%  & 10\%  &       & 30\%  & 20\%  & 20\%  &       & 20\%  & 10\%  & 20\% \\
Level 3 & 0\%   & 10\%  & 10\%  &       & 10\%  & 0\%   & 20\%  &       & 10\%  & 30\%  & 10\%  &       & 30\%  & 0\%   & 0\% \\
Level 4 & 0\%   & 0\%   & 0\%   &       & 0\%   & 10\%  & 0\%   &       & 0\%   & 0\%   & 10\%  &       & 0\%   & 0\%   & 0\% \\
Level 5 & 0\%   & 0\%   & 0\%   &       & 10\%  & 10\%  & 0\%   &       & 0\%   & 10\%  & 10\%  &       & 0\%   & 10\%  & 0\% \\
Level 6 & 0\%   & 0\%   & 0\%   &       & 0\%   & 0\%   & 0\%   &       & 0\%   & 0\%   & 0\%   &       & 0\%   & 0\%   & 10\% \\
Level 7 & 0\%   & 0\%   & 0\%   &       & 10\%  & 0\%   & 0\%   &       & 0\%   & 0\%   & 0\%   &       & 0\%   & 0\%   & 0\% \\
Level 8 & 0\%   & 0\%   & 10\%  &       & 0\%   & 0\%   & 0\%   &       & 0\%   & 0\%   & 0\%   &       & 0\%   & 0\%   & 0\% \\
Level 9 & 0\%   & 10\%  & 0\%   &       & 20\%  & 0\%   & 0\%   &       & 0\%   & 0\%   & 0\%   &       & 0\%   & 0\%   & 0\% \\
Level 10 & 10\%  & 0\%   & 10\%  &       & 0\%   & 0\%   & 0\%   &       & 0\%   & 0\%   & 0\%   &       & 0\%   & 0\%   & 0\% \\
\bottomrule
\end{tabular}%

	}
\end{center}
\end{table*}

\subsection{What is the second most important predictor?}

\begin{table*}[t]
\begin{center}
	\caption{The percentage that a given level appeared in the subset containing two levels selected by the \texttt{BE} algorithm.}\label{t3}
	\resizebox{1.0\textwidth}{!}{
	%
\begin{tabular}{lccccccccccccccc}
      &       &       &       &       &       &       &       &       &       &       &       &       &       &       &  \\
\cmidrule{2-16}      & \multicolumn{7}{c}{US Data}                           &       & \multicolumn{7}{c}{Nordic Data} \\
\cmidrule{2-8}\cmidrule{10-16}      & \multicolumn{3}{c}{DeepLOB} &       & \multicolumn{3}{c}{TABL} &       & \multicolumn{3}{c}{DeepLOB} &       & \multicolumn{3}{c}{TABL} \\
\cmidrule{2-4}\cmidrule{6-8}\cmidrule{10-12}\cmidrule{14-16}$H$   & 10    & 20    & 50    &       & 10    & 20    & 50    &       & 10    & 20    & 50    &       & 10    & 20    & 50 \\
\cmidrule{1-4}\cmidrule{6-8}\cmidrule{10-12}\cmidrule{14-16}Level 1 & 100\% & 100\% & 100\% &       & 100\% & 100\% & 100\% &       & 100\% & 100\% & 100\% &       & 80\%  & 100\% & 100\% \\
Level 2 & 10\%  & 20\%  & 15\%  &       & 15\%  & 35\%  & 25\%  &       & 5\%   & 10\%  & 30\%  &       & 35\%  & 55\%  & 50\% \\
Level 3 & 0\%   & 20\%  & 25\%  &       & 20\%  & 20\%  & 15\%  &       & 5\%   & 5\%   & 10\%  &       & 10\%  & 5\%   & 20\% \\
Level 4 & 10\%  & 0\%   & 5\%   &       & 5\%   & 20\%  & 25\%  &       & 15\%  & 10\%  & 5\%   &       & 20\%  & 5\%   & 10\% \\
Level 5 & 10\%  & 5\%   & 10\%  &       & 0\%   & 0\%   & 0\%   &       & 10\%  & 5\%   & 20\%  &       & 10\%  & 0\%   & 5\% \\
Level 6 & 5\%   & 15\%  & 5\%   &       & 20\%  & 10\%  & 15\%  &       & 5\%   & 15\%  & 5\%   &       & 10\%  & 5\%   & 10\% \\
Level 7 & 15\%  & 10\%  & 20\%  &       & 15\%  & 5\%   & 5\%   &       & 25\%  & 20\%  & 5\%   &       & 15\%  & 5\%   & 0\% \\
Level 8 & 20\%  & 15\%  & 5\%   &       & 5\%   & 5\%   & 0\%   &       & 15\%  & 10\%  & 20\%  &       & 10\%  & 15\%  & 0\% \\
Level 9 & 25\%  & 15\%  & 0\%   &       & 15\%  & 0\%   & 0\%   &       & 10\%  & 20\%  & 0\%   &       & 5\%   & 5\%   & 0\% \\
Level 10 & 5\%   & 0\%   & 15\%  &       & 5\%   & 5\%   & 15\%  &       & 10\%  & 5\%   & 5\%   &       & 5\%   & 5\%   & 5\% \\
\bottomrule
\end{tabular}%

	}
\end{center}
\end{table*}

\begin{table*}[t]
\centering
	\caption{The average percentage of appearance in a 2-element subset selected by the \texttt{BE} algorithm over all prediction horizons, neural network models and datasets, i.e., average over all columns in Table \ref{t3}.}
	\label{t4}
\resizebox{0.8\textwidth}{!}{%
\begin{tabular}{@{}cccccccccc@{}}
\toprule
Level 1 & Level 2 & Level 3 & Level 4 & Level 5 & Level 6 & Level 7 & Level 8 & Level 9 & Level 10 \\ \midrule
98.33\% & 25.42\% & 12.92\% & 10.83\% & 6.25\%  & 10.00\% & 11.67\% & 10.00\% & 7.92\%  & 6.67\%   \\ \bottomrule
\end{tabular}%
}
\end{table*}

Given that the quote data from level 1 are the most important source of information in the order-book, which level comes second? To answer this question, we looked at the percentage that a given level appeared in the subsets selected by \texttt{BE} having a cardinality of 2. Due to the sequential removal nature of \texttt{BE}, we know for certain that the numbers for level 1 are similar or higher than those appear in Table \ref{t1}. But can we observe a similar phenomenon for any other level when \texttt{BE} is allowed to form a subset of two levels?

\begin{figure*}[!h]
        \centering
        \includegraphics[width=0.6\textwidth]{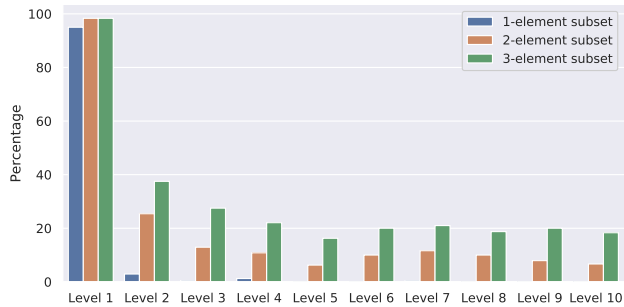}
	\caption{The average percentage of appearance of each level in the 1-element, 2-element and 3-element subsets selected by the \texttt{BE} algorithm. }\label{f2}
\end{figure*}

The statistics for this inquiry are shown in Table \ref{t3} for each dataset, each classifier, and each prediction horizon. Table \ref{t4} shows the average percentage of appearance (over all datasets, classifiers, and prediction horizons) in a 2-element subset chosen by \texttt{BE}, i.e., the average number for every row in Table \ref{t3}. The numbers indicate that the second highest record comes from level 2 at 25.42\%. By following the same procedure and computing the average percentage of each level appearing in the 1-element, 2-element and 3-element subsets yielded by \texttt{BE} algorithm, we obtained Figure \ref{f2}, which gives us an overall insights into the importance of each level assessed by \texttt{BE} algorithm.

\begin{table*}[h]
\centering
	\caption{The average percentage of appearance in the final solution outputed by the \texttt{BPSO} algorithm over all prediction horizons, neural network models and datasets, i.e., average over all columns in Table \ref{t2}}
	\label{t5}
\resizebox{0.8\textwidth}{!}{%
\begin{tabular}{@{}cccccccccc@{}}
\toprule
Level 1 & Level 2 & Level 3 & Level 4 & Level 5 & Level 6 & Level 7 & Level 8 & Level 9 & Level 10 \\ \midrule
75.83\% & 16.67\% & 10.83\% & 1.67\%  & 4.17\%  & 0.83\%  & 0.83\%  & 0.83\%  & 2.50\%  & 1.67\%   \\ \bottomrule
\end{tabular}%
}
\end{table*}

From Figure \ref{f2}, we can see that the top three quote levels of the LOB are also ranked in the same order according to the frequency of being selected by \texttt{BE}. To find potential consensus between \texttt{BE} and \texttt{BPSO}, we also computed the average percentage of appearance in the final solution of \texttt{BPSO} for each level, which is shown in Table \ref{t5}. From Figure \ref{f2} and Table \ref{t5}, we can see that there is indeed a consensus between \texttt{BE} and \texttt{BPSO} about the ranking between the top three levels while there is no consistent ranking for levels beyond the top three.

\begin{figure*}[t]
        \centering
        \includegraphics[width=0.6\textwidth]{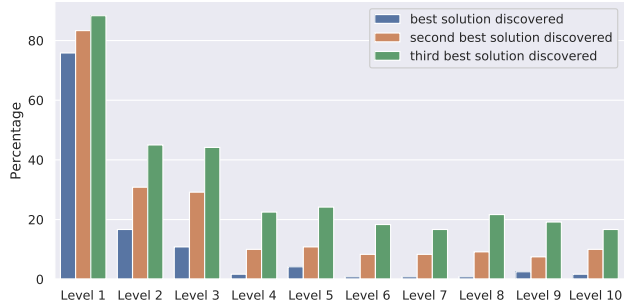}
	\caption{The average percentage of appearance of each level in the best, second best and third best solution discovered by \texttt{BPSO} algorithm during optimization process. The statistics are aggregated from both US and Nordic data for all prediction horizons}\label{f3}
\end{figure*}

For \texttt{BPSO}, we cannot specify the size of the selected subset, thus, we cannot analyze the appearance of a given level with a fixed subset cardinality as in \texttt{BE}. However, for \texttt{BPSO}, we can also look at the average percentage of appearance of each level in the second and third best solutions discovered by the swarm during the course of optimization. The statistics, aggregated from both US and Nordic data in all prediction horizons, are shown in Figure \ref{f3}, which also confirms the ranking between the top three levels of the LOB suggested by \texttt{BE}.

\subsection{Are quotes beyond the best levels needed for good performance?}

\begin{table}[h]
\centering
	\caption{Model performance reported as the mean and standard deviation of the F1 score in \%. The first line (\texttt{Baseline}) shows the performances of models that were trained using data from all levels. The second line shows the performances obtained from the most important level identified with \texttt{BE}. The third line shows the performances obtained from the subset of levels selected by \texttt{BPSO}.
}
\label{t6}
\resizebox{1\textwidth}{!}
{
\begin{tabular}{lccccccc}
\cmidrule{2-8}      & \multicolumn{3}{c}{DeepLOB} &       & \multicolumn{3}{c}{TABL} \\
\cmidrule{2-4}\cmidrule{6-8}$H$   & 10    & 20    & 50    &       & 10    & 20    & 50 \\
\cmidrule{1-4}\cmidrule{6-8}      & \multicolumn{7}{c}{Panel A: US Data} \\
\texttt{Baseline} & $65.01\pm00.36$ & $66.27\pm00.12$   & $65.73\pm00.21$  &       & $64.31\pm00.28$  & $65.26\pm00.16$  & $65.38\pm00.07$ \\
\texttt{BE} & $63.66\pm02.57$  & $60.85\pm04.43$ & $64.98\pm00.12$ &       & $60.57\pm00.84$ & $63.13\pm00.57$ & $64.02\pm00.16$ \\
\texttt{BPSO} & $64.35\pm00.22$  & $65.74\pm00.09$  & $62.26\pm03.69$  &       & $62.77\pm00.87$  & $64.38\pm00.44$ & $62.94\pm03.17$ \\
      & \multicolumn{7}{c}{Panel B: Nordic Data} \\
\texttt{Baseline} & $69.92\pm01.86$ & $71.53\pm02.22$ & $71.83\pm00.49$ &       & $70.10\pm01.23$ & $71.32\pm00.89$ & $71.89\pm00.74$ \\
\texttt{BE} & $63.77\pm08.12$ & $68.24\pm06.78$ & $71.87\pm00.08$ &       & $66.84\pm02.02$ & $69.44\pm01.35$ & $71.29\pm00.72$ \\
\texttt{BPSO} & $68.53\pm01.89$ & $68.18\pm03.36$ & $69.07\pm03.27$ &       & $67.06\pm02.43$ & $70.38\pm01.12$ & $70.15\pm03.00$ \\
\bottomrule
\end{tabular}%

}
\end{table}

In the previous analysis, we have identified that the top quotes carry the most important information for the prediction of mid-price movements. However, we have not looked at the prediction performances achieved by using the subsets of information selected by the feature selection algorithms. Table \ref{t6} shows F1 scores measured on the test set under different scenarios. The first line (\texttt{Baseline}) of each panel shows the performances obtained by training the predictors using data from all levels. The second line of each panel shows the performances obtained from the most important level identified by \texttt{BE}. The third line shows the performances obtained from the subset of data selected by \texttt{BPSO}.

It is clear from both panels that there are performance degradations using \texttt{BE} and \texttt{BPSO} in almost all configurations. More specifically, when using the subsets of data, i.e. the most informative LOB levels, selected by \texttt{BE}, we observed on average $3.76\%$ decline in performance for the US stocks, and $3.58\%$ for the Nordic stocks. Two conclusions can be drawn: on one hand, the use of the most informative LOB level data, which, according to Table \ref{t1}, most often comes from the top of the book, provides surprisingly good performance; on the other hand, the quote data deeper in the order-book provides extra performance gains. For the levels that maximize the fitness function within the training procedure with \texttt{BPSO}, the average reductions are $2.43\%$ and $3.10\%$ for US and Nordic stocks, respectively. This means that the the optimal selection of the set of LOB levels in the training phase does not lead to the superior out-of-sample performance.

Are these a few percents differences substantial or marginal? To answer this question, one needs to judge whether improvements can outweigh the extra computational cost, which depends on the specific application under consideration. Nevertheless, based on our analysis, one can be assured that there is indeed useful information contained in the levels beyond the top quotes in the LOB.

\section{Conclusion}\label{conclusion}
The major part of the existing literature uses only the top-quotes of the LOB data, which leads us to question about importance of the top quotes in comparison with quotes of lower priority. In addition, it is important to question whether the multi-level order book data can bring extra benefits compared to using only the best level data. In this paper, we answered these questions in the context of stock price prediction. Since the financial market mechanisms can be highly non-linear, we employed state-of-the-art neural network predictors in our analysis, in contrast to existing works that only employed simple linear models. The analysis was conducted using the LOB data obtained from two different markets, namely the US and the Nordic markets. To study the informational content of different LOB levels, two feature selection methods were used, namely Backward Selection (\texttt{BE}) and Binary Particle Swamp Optimization (\texttt{BPSO}). For additional robustness, different prediction horizons were used in our analysis.

Our analysis indicates that the best bid and ask levels are systematically identified not only as the most informative levels in the order books, but also to carry most of the information needed for good prediction performance. More specifically, in terms of the models' predictive power (performance), the inclusion of multiple levels from the limit order books improves the performance by around 3.5\%. Thus, even if the top-of-the-book levels contain most of the relevant information, to maximize the performance one should use all data across all the levels. In addition, the informativeness of the levels clearly decreases from the first to the fourth level while the rest of the levels are approximately equally important.

\bibliographystyle{plainnat}
\bibliography{reference}
\end{document}